\documentclass[referee]{aa}

\usepackage{ulem}
\usepackage{natbib}
\usepackage{epsfig}
\usepackage{expdlist}


\usepackage{color}
\definecolor{red}{rgb}{0.7,0,0}
\definecolor{blue}{rgb}{0,0,0.7}

\def\refe#1{\textcolor{black}{{#1}}}
\def\refee#1{\textcolor{black}{{#1}}}

\usepackage{txfonts}

\begin{document}

\title{A possible imprint of quasi-periodic oscillations in the X-ray spectra of black hole binaries}

\author{P. Varniere\inst{1}
\and R. Mignon-Risse\inst{1}
\and J. Rodriguez\inst{2}} 

\institute{APC, AstroParticule et Cosmologie, Universit\'e Paris Diderot, CNRS/IN2P3, 
CEA/Irfu, Observatoire de Paris, Sorbonne Paris Cit\'e, 
10, rue Alice Domon et L\'eonie Duquet, 75205 Paris Cedex 13, France 
varniere@apc.univ-paris7.fr
\and Laboratoire AIM, CEA/IRFU-CNRS/INSU-Universit\'e Paris Diderot, 
CEA DSM/IRFU/SAp,  F-91191 Gif-sur-Yvette, France.}

\authorrunning{Varniere, Mignon-Risse \& Rodriguez}
\titlerunning{spectral QPOs}

\date{Received <date> /
Accepted <date>}
\abstract{While nobody would deny the presence of Quasi-Periodic Oscillations in the power density spectrum of Black hole binaries nor their importance
 in the understanding of the mechanisms powering the X-ray emissions, the possible impact on the \refe{time-averaged} disk energy spectrum 
 from the phenomenon responsible for the Quasi-Periodic Oscillations is largely ignored in models of sources emission.}
{Here we investigate the potential impact of such structure on the resultant energy spectrum.}
{Using data from the well documented outbursts of XTE J$1550$-$564$ we looked at possible hints that the 
presence of Quasi-Periodic Oscillations actually impacts the energy spectrum emitted by the source. In 
particular we look at the evolution of the relation between the inner disc radius and the inner disc temperature 
obtained from fits to the spectral data. We then test this further by developing a simple model 
 in order to simulate spectra of a disk with a structure mimicking an increasing in strength Quasi-Periodic Oscillation and compare the 
 simulated results to those obtained from real data.}
{We detect a similar departure in the inner radius - inner temperature 
 curve coming from the standard fit of our simulated observations as is seen in 
 XTE J$1550$-$564$ data.
We interpret our results as evidence that the structure at the origin of the Quasi-Periodic Oscillation impact the energy spectrum.}
{Furthermore, in states with a significant disk emission the inaccuracy of  the determination of the disk parameters
increases with the strength of the Quasi-Periodic Oscillation, an increase which then renders the value given by the fit 
unreliable for strong Quasi-Periodic Oscillation.} 


\maketitle

\section{introduction}
  
 When looking at the Power Density Spectrum (PDS) of microquasars the most striking features are the presence of 
 narrow peaks, called Quasi-Periodic Oscillations (QPO). 
 These Low-Frequency ($<30$Hz, LF) and High-Frequency ($>40$Hz, HF) QPOs 
 contain a significant part of the source rapid variability. Some LFQPOs can have an amplitude of up to $30$\%.
Several distinct models exist to describe them, and many imply a warm/hot structure orbiting the disk causing the X-ray modulation. 
  Such structures include axisymmetric~\citep[see e.g.][]{B04, S06b, Vin14},  or precessing tori~\citep{S06c, I09},
   hotspots~\citep{K92, S04, TV06, P13} or spirals~\citep{TP99,VR02}.

  Nevertheless, when analyzing the sources energy spectra the disk is mostly
  considered as relatively homogeneous and with a smooth,  monotonic temperature profile.
  If such a featureless disk may be a good representation for the so-called thermal states during which
  the energy spectra resemble ``pure" blackbody and the PDS associated show little or no variability, with no or weak QPO, 
  this model is not appropriate to describe states with prominent QPOs. 
  Here we aim to explore if the structure at the origin of QPOs indeed has a measurable impact on the energy spectrum and its fits.
        
  The shape and the existence of more than one component in the  X-ray spectra makes it 
  hard to look at a direct impact from the structure at the origin of the QPO. Nevertheless, we can look at a possible impact 
  of the presence of QPOs on  the spectral fit parameters and the correlations between them.    
  Many correlations between spectral parameters and QPO parameters have been presented \citep{M99,R02,Ro02,V03} but not many studies
  have been done to investigate any change in the correlation between spectral parameters depending on the QPO presence.
  
  In section \ref{sec:obs} we  use data from \refe{two outbursts of XTE J$1550$-$564$ observed by RXTE},
   which have the advantage of being long and well observed, to see
  if some trends are indeed visible in the correlation between spectral parameters  depending on the presence of QPOs.  
 In order to test further this idea, we present in section \ref{sec:model} a simplified  model of a disk having an 
 increasingly strong QPO that we then use to compute the source's energy spectrum.
 In section \ref{sec:impact} we use this model in conjecture with the module {\tt fakeit} from XSPEC to test the impact on the energy
 spectrum of a hot structure in the disk.
 This allows us to directly measure the impact of the QPO as the difference between the
  model parameters and the resulting fits but also to compare those simulated fits results against real correlations. 
 
\section{Link between the disk parameters and QPOs}
\label{sec:obs}      

	It has been asserted early on that the properties of the LFQPOs, in particular their frequencies, are related to parameters of the disk 
	\citep{M99,Ro02,V03,VR02}  obtained through spectral fitting. 
	As there is much less data for HFQPOs, no similar study has been done yet but HFQPOs seem to be linked with the presence of 
	LFQPOs of type A or B \citep{R02}.  
	Here we are using data from XTE J$1550$-$564$ during the outbursts of $98$-$99$ and $2000$, both known 
	to harbor HFQPOs, to see if there is any difference in the behavior of the different spectral parameters depending if there is or not 
	a QPO of any types observed. \\
	
\subsection{Observations and Data Reduction}	

XTE J1554$-$564 has undergone 5 different outbursts during the lifetime of 
{\it{RXTE}}. The last three occurred in 2001, 2002, and 2003 are considered as ``failed" since the source 
did not show any spectral transition to soft flavored (i.e. disk) states. We therefore consider here only the 
first two outbursts that also are the two brightest. 
They respectively occurred from 1998,  Sept 6, until roughly the beginning of May 1999 for the discovery one  and 
from 2000, April 10 to 2000, July 16 for the second and fainter outburst. Both outbursts 
have been extensively described in the literature \citep[e.g][]{S00,R02,Miller01,Rodriguez03,Rodriguez04}.
In the following study we consider only the observations showing the presence of LFQPO and/or HFQPO as reported 
in these articles. The QPO parameters are taken from \citet{R02,Miller01,Rodriguez04}. 
Since the spectral calibration of the Proportional Counter Array (PCA) has significantly evolved since 
the publication of these articles, we re-reduced and re-analyzed the spectral data in order to present 
the most accurate spectral parameters \refee{but we kept the parameters. In order to compare with the previous 
results \citep[e.g][]{R02} we kept  the distance ($D=6$ kpc) and inclination ($70^\circ$) that were used in 
these earlier papers.  The distance and inclination we use are only a scaling factor applied to the radius 
(see full expression if eq \ref{eq:T}). As long as we use the same for the observations
and the simulated observations we can compare the results.}

The PCA data were reduced with the {\tt{HEASOFT}} v6.16 suite. Good time intervals were defined as 
intervals with a satellite elevation angle above the Earth greater than 10$^\circ$,  an offset between the pointing and  
the source direction less than $0.02^\circ$, and recommended limitations on proportional counter units (PCU) 
potential breakdowns. 
Spectra were then extracted from the top layer of PCU 2. Background spectra were obtained with the bright 
background model and generated with {\tt{pcabackest}}, while response files were generated through 
{\tt{pcarsp}}. 0.8\% systematic errors were added to all spectral channels before fitting. 

The PCA spectra were fitted with {\tt{XSPEC}} v12.6.2, between 3 and 30 keV. As we were interested in basic parameters we only considered 
simple spectral models. All spectra are well represented with an absorbed ({\tt{phabs}}) disk ({\tt{diskbb}}) plus 
power law ({\tt{powerlaw}}) model. \refe{The value of the high spectral boundary of our fits did not allow us to significantly constrain the  
potential presence and properties of a reflection bump. We nevertheless note that  a Gaussian at 6.5~keV representing 
fluorescent emission of iron, usually taken as a signature for reflection,  was, in a few cases, added to the model to obtain a good fit. }
The absorption was fixed to N$_{\mathrm{H}}=0.8\times10^{22}$~cm$^{-2}$ \citep{Miller03}.
\newline

\refe{The spectral fit gives access to the slope of the power-law (i.e. the photon index $\Gamma$), its normalization, 
the temperature of the inner edge of the disk and the disk normalization. The latter is related to the inner apparent disk radius $R_{in}$ 
following ${\mathrm{Norm}}=((R_{in}/km)/(D/(10kpc)))^2 \cos i$ with $D$ the distance to the source and $i$ the viewing angle of the disk.  
It is known that the apparent radius relates to the real inner radius through a "hardening factor" $f$ \citep{Shimura95}. 
Here we choose to keep that $R_{in}$ and not add a correction factor as we are interested in the raw output
of the fit that we then intend to compare with our simulated observations.}

\subsection{Departure from $r_{in}$-$T_{in}$  correlation}
	
\begin{figure}[ht!]
\centering      
\includegraphics[width=8cm]{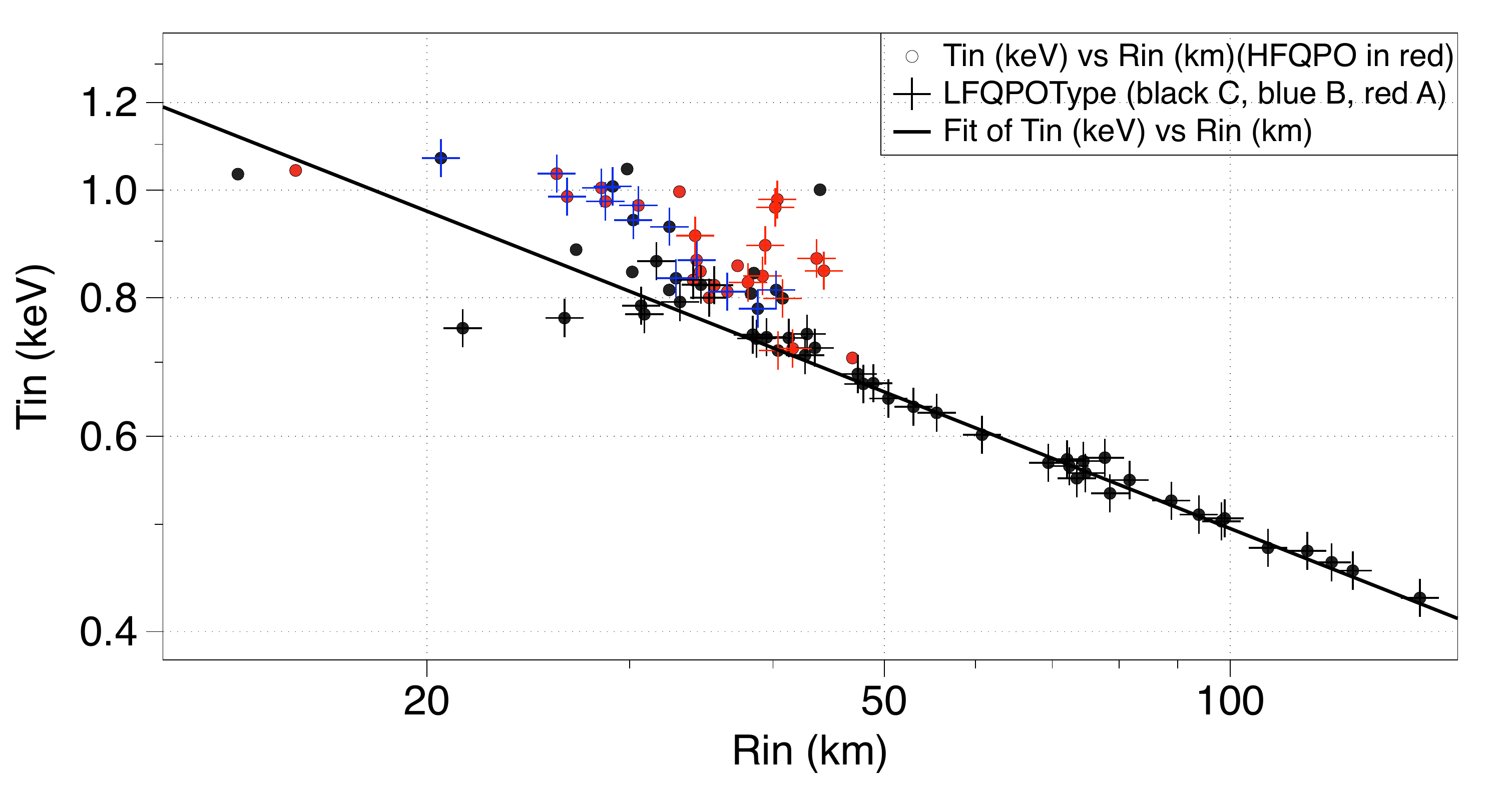}
 \caption{Correlation between the inner edge position and inner edge temperature \refe{as directly given by the spectral fits} for the outburst of $98$-$99$ and $2000$ 
 of  XTE J$1550$-$564$. Red dots represent observations with HFQPOs detected while there is none in the black dots. 
 The crosses represent the type of LFQPO, black for the common type C, blue for type B and red for type A. Timing data from \citet{R02,Ro02}.}
 \label{varniere2:fig1} 
 \end{figure}

    In Fig. \ref{varniere2:fig1} we represent the evolution of the disk parameters $r_{in}$ vs. $T_{in}$ obtained from the spectral fits to the data.  
        Two behavior emerge from this plot, at large radius the majority of points 
    narrowly follow a power-law, while there is an obvious departure as the inner edge of the disk gets smaller than $50$~km.
    Interestingly this departure occurs when 
    HFQPOs or similarly type A/B LFQPOs are observed\footnote{As a side note we are looking to see if the few points that 
    depart from the correlation without a published HFQPO frequency have or not some high frequency structure fine enough to be a HFQPO.}. 
    It is also interesting to note that this departure is always in the same direction, namely, at similar inner radius, we get a hotter disk in the case of 
    a detected HFQPO/type A/B LFQPO than in the standard case with a type C (hard state).
     This may be compatible with the HFQPOs  \refe{being some kind of hot structure} in the disk. 
    
    At this point, one has to wonder if, as all those data points come from the spectral fitting with a smooth monotonic disk, it is possible that the
    departure from the correlation could be rooted from the presence of a QPO/warm structure which causes the fit {\it to fail and gives inaccurate values}.
      In order to test this assumption, that {\bf in the presence of a localized hot structure in the disk spectral fit using a smooth temperature
      profile will give inaccurate value},  we will perform fits of a disk with such structure but in a controlled, hence simulated, manner and 
      compare the results of the fit with the real values.

\section{Simple model for the temperature profile}
\label{sec:model}

	QPO models often connect the signal modulations to  the presence of structures embedded in the disk.
	Our main concern here is not the origin of the structure, but the consequences on the emission and if it ever needs to be taken
	into account when doing a \refe{spectral fit}. This means that, rather than taking full magnetohydrodynamic (MHD) 
        simulations of the different models proposed to explain QPOs we are interested in, we decided to create a simple, 
        analytical, model in order to test in a cleaner and easier way,  its impact.   
        Indeed, in a full fluid simulation changing one parameter in the initial condition can have repercussions on several observable parameters 
         and therefore  it is harder to study the different effects separately.\\
     
     	Here we are using a similar perturbative approach as in \citet{VB05} and take a regular {\tt diskbb} disk model with  $T_0(r) \propto r^{-3/4}$ 
	on which we add a component mimicking the structure at the origin of the QPOs 
	that depends on  radius and azimuthal angle  as $T_1(r,\varphi) = h(r)\ . \ s(r-r_s,\varphi)$.  	
	For simplicity we choose to decompose $T_1$ as a height function $h$ that depends only on $r$ and a shape function $s$ 
      which is finite only near the disk structure we are studying which is in turn defined by $r_s=r_s(t,\varphi)$.	
	Also for simplicity we take the shape function to be gaussian and the height function to be  
		related to the equilibrium temperature $T_0(r)$.  \\

	Using this structure we can take into account a variety of shapes mimicking a variety of models, for example:
	
\noindent \refe{- a torus  with $r_s(t,\varphi)=$cte$\times \sin(\nu t)$ and $\nu$ the oscillating frequency of the torus,}\\	
\noindent \refe{- a hotspot  with $r_s(t,\varphi)=$ cte for the hotspot azimuthal size and zero otherwise.}\\
	\refe{In the general case this constant, which defines the position of the structure in the disk, is called $r_c$, the corotation radius
	of the structure. \refee{In the case of non-axisymmetrical structures such as a blob 
	$\Omega(r_c)$, the frequency at which the structure is rotating is also the frequency  at which the flux
	will be modulated, hence the QPO frequency.}}
\newline

    This provides a simple but useful framework to model a disk with added perturbative structures. Within this framework the perturbed temperature reads
 \begin{eqnarray}
   \label{eq:T}
  T(r,\varphi) &=& T_0(r)  \times\left[  1 + \gamma \ 
 \mathrm{exp}\left(-\frac{1}{2}\left(\frac{r-r_s(\varphi)}{\delta \,r_c}\right)^2\right)\right]^2 \\ \nonumber
 \end{eqnarray}
   where $r_c(t)$ is the position of the temperature maximum in the disk (the center of the torus or hotspot). 
       $\delta$ parametrizes the radial extent of the structure while $\gamma$  is the maximum amplitude of the perturbation.
   \refe{In the general case}  $r_c$  is allowed to be a function of time \refe{to take into account a change in frequency of the modulation,
   but here we are considering only the case where the QPO frequency is stable during one \lq simulated observation\rq\ hence 
   keeping $r_c$ a constant. \refee{Indeed, in the blob model the frequency of the modulation is $\Omega(r_c)$.}
   In that case, the width of the peak in the PDS represents the radial extent of the structure, here parametrized by $\delta$.}
   
   \refe{These last two parameters, ($\gamma, \delta$) are the ones we will be using to mimic a growing flux modulation.}
   Indeed, using similar models we are able to reproduce several timing observables such as, for example, the rms  amplitude of 
   QPOs as shown in \cite{VB05} for a  pseudo-newtonian potential and \cite{VV16} in general relativy, 
    hence validating this simple model as a representation of a disk giving rise to a flux modulation, hence a QPO.
    All the simulations presented in Table \ref{tab:fit} 
   represent an evolving flux modulation from zero to about $\sim20$\% hence well inside the observed limits. \\

Using these parametrized temperature profiles we then can compute the emitted thermal flux from 
the entire disk \refe{considering it made of multiple blackbody components} in a similar manner  as done in the widely used {\tt{XSPEC}} 
model  diskbb for a monotonic profile:       
 \begin{eqnarray}
    \label{eq:f}
f(E) = \frac{r_{in}^2}{D^2} \cos{i} \frac{2E^3}{c^2h^3} \int_{0}^{2\pi}\int_{1}^{\infty} \frac{r_\star}{e^{\frac{E}{k_b{\bf T}(r_\star,\phi)}}-1}   dr_\star d\phi  
 \end{eqnarray}   

where \refe{$r_\star$ is the radius in units of $r_{in}$} the inner disk radius, $k_b$ is Boltzmann's constant, $h$ is Planck's constant and $c$ the speed of light. The only difference with the diskbb model is that \refe{our $T(r,\phi)$ is a non-monotonic function of $r$ and also of $\phi$}, hence we have to keep both integrals.  
We then created an XSPEC model of disk with this structure (which we named
 {\tt diskblob} for the case of an elongated hotspot)  which will be made available once optimized.

 This allows two things: first we can fit observations with our non-monotonic disk profile and second, 
 using the procedure {\tt fakeit} in XSPEC we can create synthetic spectra of a power-law plus the disk taking into account  the presence of a warm/hot structure. Using the simulated spectra we can then test if the presence of a structure causing a QPO in the PDS would also have a detectable 
 impact on the energy spectrum.
\refe{Using  {\tt fakeit} in XSPEC gives us several limitations as we cannot use time dependent temperature profiles, hence we cannot take into
account the doppler boosting on the structure. Both of those limitations can be addressed.}
 \refe{Indeed, the simulated spectra are made as observations of $3000$ seconds, hence are averaged over several thousands of QPO periods. 
 For each \lq observation\rq\ we took a constant QPO frequency (namely $r_c$ is constant) so the temperature profile for the case of an elongated blob 
 is conserved by rotation at the QPO frequency, which allows us to use the time-average spectrum\footnote{
 \refee{We also compared an extended blob with a fully axi-symmetrical (ring) structure with the same temperature profile and found the results to be globally consistent. Indeed the average profile of a blob of finite azimuthal size would be smaller than the axi-symmetric (azimuth of $2\pi$) version of the same blob. This comforted us that a hotter structure present in the disk will indeed have an effect on the overall energy-spectrum.}}. }
 \refe{Concerning the impact of the Doppler boosting, this effect is greatly diminished in the case of a face-on disk so the calculation done here
  will be valid for a face-on disk and represent a lower limit for more edge-on disks as the Doppler effect boosts the emission of the hotspot, hence increasing its impact. Here we are interested to see if this lower limit is already detectable.}

\section{Impact on the energy spectrum fitting}
\label{sec:impact}

	In order to see if the presence of a QPO has any impact on the energy spectrum  we computed severals  synthetic spectra.
	We used the parameters of XTE J$1550$-$564$ which is 
	extensively observed and against which we could then test our	results. 
	For each simulation we will use the procedure {\tt fakeit} on an absorbed ({\tt{phabs}}) disk ({\tt{diskblob}}) plus 
	power law ({\tt{powerlaw}}) model. { \tt{diskblob}} will allow us to mimic a  disk with a slowly increasing hot structure
	aiming to reproduce the time evolution effect of a growing QPO.
	The latter is introduced in the model through two parameters (in that case $\gamma$ and $\delta$ only).
	We then fitted the synthetic spectra with XSPEC following the standard procedure with {\tt diskbb} and {\tt powerlaw}.  \\

           Using the model presented in Eq.\ref{eq:T} $\gamma$ represents the amplitude/strength of the instability causing the structure while $\delta$
            parametrized its radial extent, which can, in turn, relate to the FWHM of the QPO we are modeling. The value of
            $\gamma$ and $\delta$ in Table \ref{tab:fit} are coherent with the QPO features one would expect during an outburst 
            \citep[as seen in the general case in][then for a particular outburst in Varniere \& Vincent 2016b]{VB05}.
            \refe{Here we took an  \lq origin point\rq\  from the curve on Fig.\ref{varniere2:fig1}.  at  $(r_{in}, T_{in}) = (45, 0.7)$ where
            $T_{in}$ is the normalization of $T_0$ in Eq.\ref{eq:T}. We chose this point as it is just before we start seeing a departure in the
            correlation.}
            This means that, in all our simulations,  
            our disk has an inner edge at $r_{in}=45$km and  an associated inner edge  temperature of $0.7$keV. 
            This allow us to see
            the discrepancy between the \lq real\rq\ value and the value given by the fit, measured by $\Delta_X = (X^{real}-X^{fit})/X^{real})$.
            This discrepancy is a measure of the accuracy of the fit and its reliability.
           In all the sets in Table \ref{tab:fit} $(r_{in}, T_{in}) = (45, 0.7)$ and only the couple $(\gamma, \delta)$ is modified.

\begin{table}[htbp]
\centering
\begin{tabular}{|cc|cc|}
$\gamma$ &$\delta$ & $T_{in}^{f\refe{it}}$ & $R_{in}^{f\refe{it}}$ \\
\hline
\hline
0   &	0.05  &	0.71  &	41.4 \\
0.1   &	0.05   &	0.73   &	40.1 \\
0.1   &	0.1   &	0.69   &	48.2 \\
0.2   &	0.1   &	0.70   &	44.2 \\
0.3   &	0.1   &	0.74   &	40.9 \\
0.4   &	0.1   &	0.79   &	36.8 \\
0.4   &	0.15   &	0.80   &	37.9 \\
0.45   &	0.1   &	0.81   &	35.5 \\
0.45   &	0.15   &	0.82   &	36.1 \\
0.45   &	0.18   &	0.86   &	34.1 \\
0.5   &	0.15   &	0.84   &	35.4 \\
0.5   &	0.18   &	0.87   &	33.9 \\
0.5   &	0.2   &	0.89   &	33.8 \\
0.7   &	0.1   &	0.92   &	28.6 \\
0.7   &	0.15   &	0.98   &	27.9 \\
0.7   &	0.18   &	0.99   &	28.6 \\
0.7   &	0.2   &	1.02   &	27.5 \\
0.9   &	0.18   &	1.14   &	24.0 \\
0.9   &	0.2   &	1.16   &	24.8 \\  						      
1   &	0.18   &	1.22   &	22.8 \\
\hline
\end{tabular}
\caption{Hot structure parameters and the results from the fit, as a reminder for all of those the {\it real} inner edge of the disk is at $45$km and the
temperature is $0.7$keV. All those have an amplitude of modulations between zero and $20$\%, see \citet{VV16} for details on the amplitudes calculation.}
\label{tab:fit}
\end{table}
	 
	Following the evolution of the fit parameters, we see that as soon as the QPO has a non-zero amplitude there is a 
	discrepancy between the fitted value of the
	disk parameters given by the spectral fit and  the real value used to create the spectrum.

	This discrepancy grows with the QPO amplitude (represented by the parameter $\gamma$ in our 
	model):  already with a feature having  parameters ($\gamma, \delta$) =($0.4, 0.15$), which translate
 	 in a {$5$\% flux modulation
	 we get $\Delta_{T_{in}} = 14$\%  and $\Delta_{R_{in}} = 12$\%.
         It is interesting to note that the fitted value for  $r_{in}$ are almost always  smaller than the actual value of the disk we input
         while the $T_{in}$ are almost always overestimated.
         Indeed in our case of a \lq real\rq\ inner edge at $45$km we get some fit results up to about $23$km for our largest features
         which is equivalent to an rms of $20$\%.   
         \refe{While the discrepancy is always higher for the temperature, the impact on the determination of the inner radius of the disk is actually
         more important. Indeed, the minimal position of the inner edge of the disk is often used to put some limit on the spin of the black hole.
         In that case an underdetermination by almost factor of two between the  \lq real\rq\ value and the fitted one would lead to a much higher
         inferred spin. In the case here, where the inner edge of the disk is at $45$km, which is approximatively
         $3r_g$ for a $10 M_\odot$, the fit gives back a alue of $23$km representing  about $1.6r_g$, which would in turn indicate an almost maximal spin.
         As this happens when we have strong QPO, it is better to continue using only the value of the radius from states without QPOs to 
         put constraints  on the central object spin.}
 \newline        
	 
	Furthermore, when we plot the results of the  fit from the synthetic spectra together with data points from XTE J$1550$-$564$ we see,
	in  Fig. \ref{varniere2:fig2}., that the grey star representing our simulated spectra occupy the same space as the HFQPO/type B LFQPO 
	data points from XTE J$1550$-$564$ hence strengthening the link between  QPO, hot spot, and fit-difficulties. 
\begin{figure*}[ht!]
 \centering
\includegraphics[width=15cm]{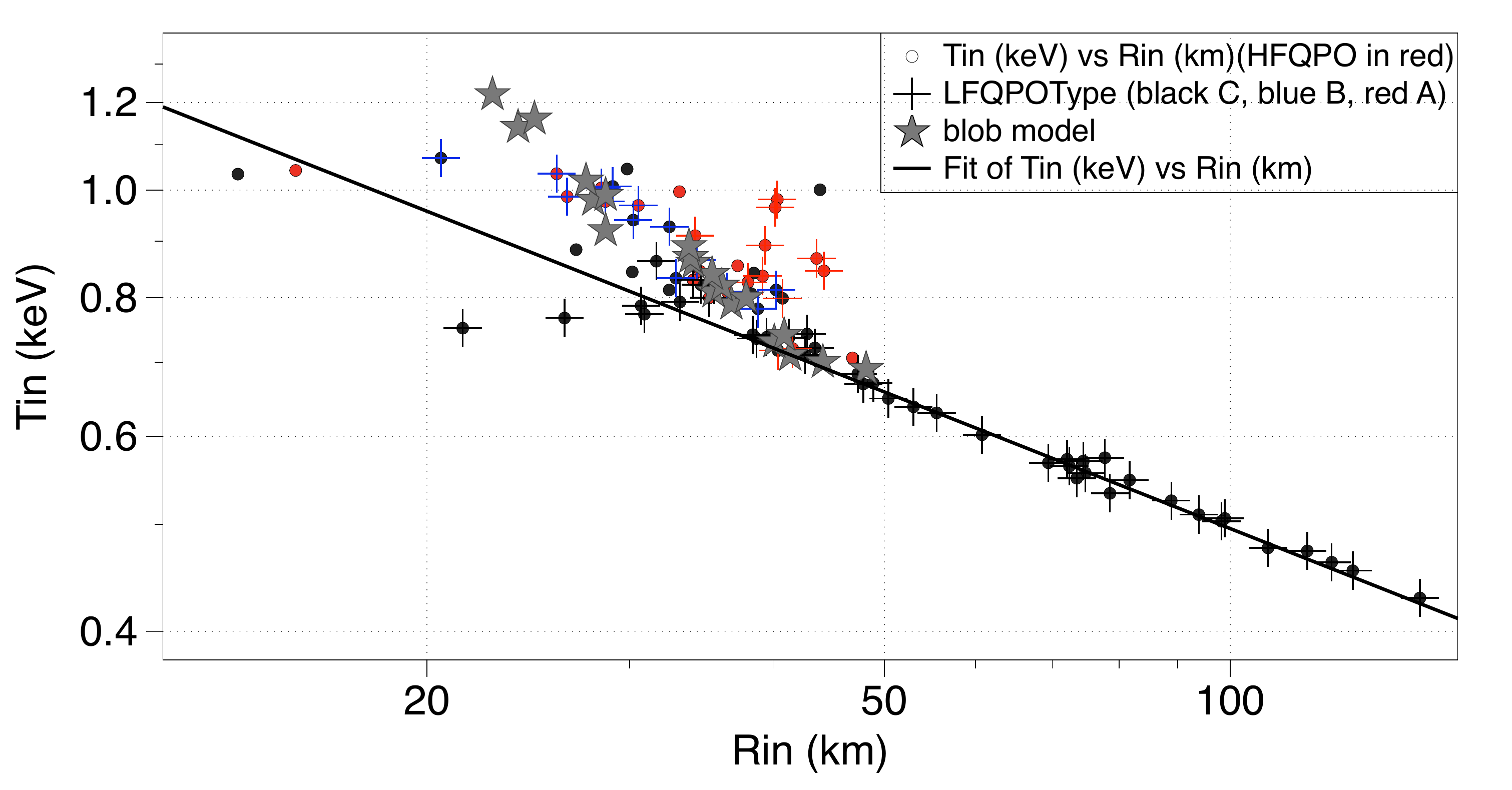}
  \caption{Correlation between the inner edge position and inner edge temperature as given by the spectral fits for the outburst of $98$-$99$ and $2000$ of  XTE J$1550$-$564$.
Red dots represent observations with HFQPOs detected while there is none in the black dots. The crosses represent the type of LFQPO, black for the common type C,
blue for type B and red for type A. The grey stars are the result of the synthetic spectra fit. Timing data from \citet{R02,Ro02}.}
  \label{varniere2:fig2}
\end{figure*}

         We see that, even with the RXTE resolution and range, neglecting the QPO in the spectral analysis can lead to large
          discrepancies. 
         As a side note, such departure from correlation could be used to detect, purely from spectral analysis, the possible presence of HFQPOs
         but would not allow to predict its parameters as we cannot disentangle the impact of  $\delta$ and $\gamma$ on the spectrum.

\section{Conclusion}

     Using a simple model to mimic the emission from a disk with a non-monotonic temperature, as has been theorized to be the case
  in the presence of QPOs, we have created synthetic spectra of a system exhibiting  an increasingly strong QPOs.    
  This allowed us to study in a clean environment the impact of a the structure at the origin of the QPO on the energy spectrum 
  and determine if we can neglect them in the spectral fit. \\
  
     First, our simulated observations are coherent with the departure from correlation seen in the $T_{in}$-$r_{in}$ diagram of XTE J$1550$-$564$ in
    presence of HFQPO and LFQPOs B or A. In the case of very small amplitude QPOs there is a negligible impact on the energy spectrum and it 
    can be ignored in the spectral fit. 
    
    \refe{Nevertheless, in presence of a medium strength/impact QPO (more than  $5$\% rms amplitude)} 
    we cannot neglect the presence of the hot structure as it leads to significant 
    discrepancy between the fitted value and the physical parameters. 
    Especially those errors tend to almost exclusively give a smaller inner radius and higher inner temperature.	
    Therefore there is a need to improve the disk fitting by taking into account the structures at the origin of the QPOs if we want to constrain the 
    disk parameters in their presence.

\begin{acknowledgements}
The author thanks the anonymous referee that helped clarify the paper to this final form. We acknowledge the financial support of the UnivEarthS Labex program at Sorbonne Paris Cite (ANR-10-LABX-0023 and ANR-11-IDEX-0005-02).
\end{acknowledgements}


\end{document}